# Implementing the Three-Stage Quantum Cryptography Protocol

Priya Sivakumar

**Abstract**:

We present simple implementations of Kak's three-stage quantum cryptography protocol. The case where the transformation is applied to more than one qubit at one time is also examined.

**Introduction:**

Quantum cryptography protocols are either hybrid, in which information between Alice and Bob is partly exchanged over a quantum channel and partly over a classical channel (e.g. [1,2]), or purely quantum, as in Kak's three-stage protocol [3], although when using certificates this also become hybrid [4].

The original three-stage protocol paper suggested the use of rotation operators. Here we propose two simpler implementations of the protocol, and also discuss the use of transformations that apply on several qubits simultaneously.

**The Protocol and its Implementations**

In the three-stage protocol, the security is based on the fact that Alice and Bob, the transmitter and receiver of the information, use secret keys in the multiple exchange of the qubit. Since this protocol remains quantum in each stage, the eavesdropper cannot obtain information without disrupting the communication process.

The sequence of steps is summarized in Figure 1:

1. Alice applies the transformation $U_A$ on the secret qubit S and sends the qubit to Bob.
2. Bob applies $U_B$ on the received qubit $U_A(S)$ and sends it back to Alice.
3. Alice applies $U_A^\dagger$ on the received qubit, converting it to $U_B(S)$, and forwards it to Bob.
4. Bob applies $U_B^\dagger$ on the qubit, converting it to S.

It is an obvious requirement that the transformations $U_A$ and $U_B$ commute.

We suggest that Alice and Bob agree to the use of operators from a mutually decided set. This is only slightly more restrictive than the requirement that the transformations used be commutative, since that would also necessitate some agreement between the two parties.

We propose that the use of Pauli transformations is more convenient than the use of rotation operators, because it entails less severe precision requirements than in arbitrary rotations. Note, furthermore, that in rotation operators one must also account for the rotation caused by the turnaround of the qubit at each re-transmission, and not overlook the difficulties with maintaining fidelity in the presence of noise (e.g. [5-7]).

Since we wish to operate only on well-defined states, we need not worry about the question of the difficulty of implementing gates [8,9]. The fidelity of transmission can be checked in each of these cases by the exchange of parity information on subsets of received bits as is done in the BB84 protocol [1].

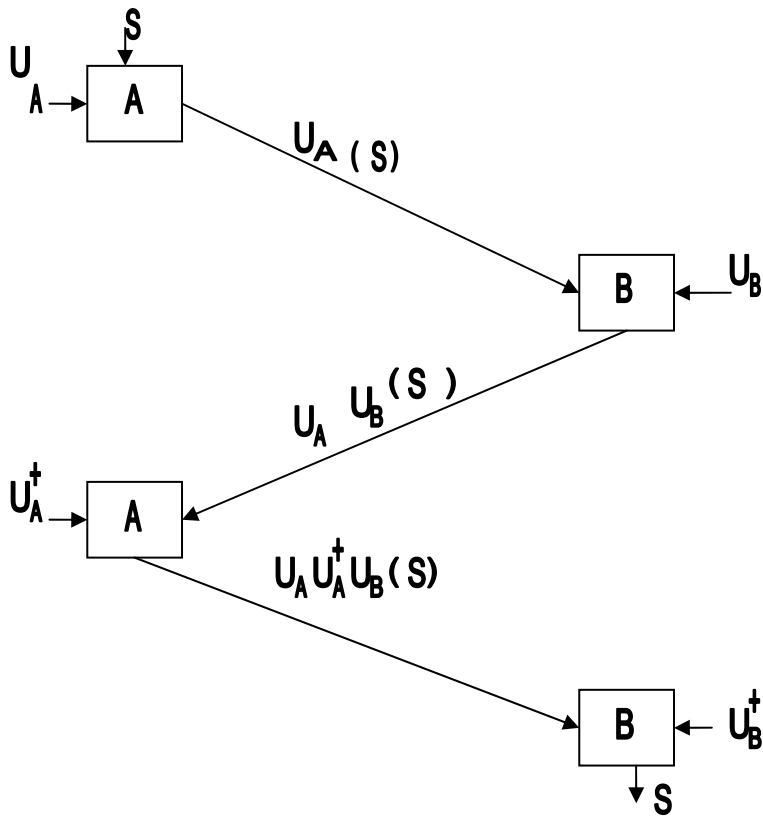

**Figure 1.** The three-stage quantum cryptography protocol

**First implementation:**

Our objective is to find a group of transformations for use in the protocol. The only condition is that the transformations should map into the |0> and |1> states with equal probability for the sake of cryptographic security. The simplest such group consists of the basic single-qubit operators I, X, Y, Z:

$$I = \begin{pmatrix} 1 & 0 \\ 0 & 1 \end{pmatrix}, X = \begin{pmatrix} 0 & 1 \\ 1 & 0 \end{pmatrix}, Y = \begin{pmatrix} 0 & -i \\ i & 0 \end{pmatrix}, \text{ and } Z = \begin{pmatrix} 1 & 0 \\ 0 & -1 \end{pmatrix} \quad (1)$$

These operators commute as global phase may be ignored.

Alice and Bob pick any of the 4 operators to operate upon S. A secret S= |0> will become a |1> with 50% probability upon the use of one of these four transformations, establishing security of this procedure.

**Second implementation:**

Let the set of transformations out of which Alice and Bob choose be K and L, which are the do-nothing and the Hadamard transformations, which also form a group:

$$K = \begin{pmatrix} 1 & 0 \\ 0 & 1 \end{pmatrix} \text{ and } L = \frac{1}{\sqrt{2}} \begin{pmatrix} 1 & 1 \\ 1 & -1 \end{pmatrix} \quad (2)$$

Once again, this provides security since there is a fifty percent chance that the initial state has been put into a superposition.

**Two qubit system**

This can be generalized easily by considering transformations on several qubits at the same time, which only requires that the transformations remain commutative. This is easily accomplished. For example, Alice and Bob may use the transformations $U_A$ and $U_B$ given below:

$$U_A = \begin{pmatrix} 1000 \\ 0100 \\ 0001 \\ 0010 \end{pmatrix} \text{ and } U_B = \begin{pmatrix} 0100 \\ 1000 \\ 0010 \\ 0001 \end{pmatrix} \quad (3)$$

As another example, for 2-qubit transformations, the protocol might be for Alice and Bob to either perform the DFT or do nothing at all. For the two-qubit case, the transformation is:

$$U_A = U_B = \frac{1}{2} \begin{bmatrix} 1 & 1 & 1 & 1 \\ 1 & i & -1 & -i \\ 1 & -1 & 1 & -1 \\ 1 & -i & -1 & i \end{bmatrix} \quad (4)$$

A third proposal is to use the four unitary 4×4 matrices that map into quarternions [10]:

$$\begin{bmatrix} 0 & 1 & 0 & 0 \\ -1 & 0 & 0 & 0 \\ 0 & 0 & 0 & 1 \\ 0 & 0 & -1 & 0 \end{bmatrix} \begin{bmatrix} 0 & 0 & 0 & -1 \\ 0 & 0 & -1 & 0 \\ 0 & 1 & 0 & 0 \\ 1 & 0 & 0 & 0 \end{bmatrix} \begin{bmatrix} 0 & 0 & -1 & 0 \\ 0 & 0 & 0 & 1 \\ 1 & 0 & 0 & 0 \\ 0 & -1 & 0 & 0 \end{bmatrix} \begin{bmatrix} 1 & 0 & 0 & 0 \\ 0 & 1 & 0 & 0 \\ 0 & 0 & 1 & 0 \\ 0 & 0 & 0 & 1 \end{bmatrix}. \quad (5)$$

They commute due to the fact that global phase may be ignored. These matrices are tensor products of the Pauli matrices of (1) and, therefore, these transformations represent the same operations as in (1) when considered in pairs.

**Conclusions**

The use of transformations that belong to a group, while ensuring that the two measurement states have equal probability, has helped find good implementations of the three-stage quantum key distribution protocol.